\documentstyle[twocolumn,aps,prl]{revtex}
\begin{document}


\maketitle


{\large\bf Comment on ``Integrability and Coherence of Hopping
between 1D Correlated Electron Systems''} \\
~\\
In a recent Letter \cite{Mila}, Mila and Poilblanc
numerically investigated the time dependence of
a non-equilibrium expectation value 
previously introduced by us \cite{usprl} in connection
with our proposal for the potential incoherence
of single particle hopping between non-Fermi liquids.
They interpret their results as indicating the
importance of integrability for the question 
of the coherence of the interchain hopping;
we believe that their results indicate the importance
of integrability for questions of {\it ergodicity},
rather than coherence.

In the context of our proposal that
the renormalization
group relevance of interliquid hopping is insufficient
to demonstrate that {\it coherent} \cite{usprl,def} interliquid
hopping occurs,
we have previously suggested that interference effects
might be the most natural probe of coherence.
In particular, one natural quantity to study 
is defined as follows:
prepare two Luttinger liquid chains in the zero interchain hopping
($t_{\perp} = 0$)
groundstate, with some inequality of particle
number between the two chains,
$\delta N (t=0)$, turn on $t_{\perp}$ at time $t=0$
and then study the expectation
value of the number difference,
$\langle \delta N (t) \rangle$; oscillations 
in $\langle \delta N (t) \rangle$ are an  interference effect and
their absence indicates incoherence \cite{usprl,advances}.

We have also introduced \cite{usprl}
the quantity $P(t)$, the 
probability of finding the system in its initial state
at time $t$ after turning on the hopping under the preparation
described above.  
This is the quantity
investigated numerically in \cite{Mila}. 
Our reason for studying $P(t)$ was 
that we wished to avoid certain pathologies
associated with studying $\langle \delta N (t) \rangle$
at low order in perturbation theory \cite{usprl,advances}.
At lowest order,
a study of $P(t)$ addresses the question of the degeneracy or
non-degeneracy of the perturbation theory in
$t_{\perp}$ which is the central question of
coherence \cite{usprl,advances}. At higher orders, or longer times,
$P(t)$ is a much more ambiguous probe of coherence than
$\langle \delta N (t) \rangle$.

As defined,
$P(t)$ is the probability to find the system in its
initial {\it microstate} at time $t$.  
Thus, in the thermodynamic limit for any reasonably
ergodic system, $P(t)$ should decay to
zero with a rate that diverges with the system
size, $L$, and recurrences in $P(t)$ will
be exponentially small in $L$ \cite{Nbetter}.
For finite systems,
it is unclear,
even in the presence of coherence,
whether one expects to be able
to see any significant oscillatory component
to $P(t)$ for reasonable $L$ and typical ergodic properties.
These difficulties imply that the apparent vanishing of the amplitude
of the oscillations in $P(t)$ 
is not sufficient to establish 
incoherence \cite{freq}.  
Unfortunately, 
it is this premise which underlies
the central conclusions of
\cite{Mila}.

We believe that the disappearance of the oscillations in $P(t)$
observed in \cite{Mila} away from integrable points
is more naturally explained by the fact
that integrable systems, with their infinite number
of conservation laws, have unusual ergodic properties
which enhance the chance of revisiting a
specific microstate.  
We note that
integrability is {\it not} a low energy, universal
property:
while the $t-J$ model studied 
in \cite{Mila} is integrable only at special points,
its low energy properties for
small $J/t$ should be the same as those of
the Hubbard model for large $U$ - an integrable model.  
More generally,
irrelevant operators which do not affect
low energy properties can spoil integrability.
Any change in behavior in $P(t)$
associated with integrability is related to
non-universal features of the coupled chain
model of \cite{Mila}.
While it seems natural that the
ergodic properties of a model could depend
on its non-universal features, particularly 
its integrability, for incoherence,
no dependence on non-universal properties 
should occur in the limit of small $t_{\perp}$.
We are therefore of the
opinion that, while the results of \cite{Mila}
may be of interest in connection
with questions of ergodicity,
they are not directly meaningful for the question of coherence \cite{final}.\\
~\\
{\large David G. Clarke$^1$ and S.~P.~Strong$^2$} \\
$^1$ Department of Physics, Princeton University, \\
Princeton, NJ, 08544-0708 \\
$^2$Institute for Advanced Study,\\
Olden Lane, Princeton, NJ, 08540


%
%

%
%

\end{document}